\documentclass[amsmath,amssymb,prc,final,showpacs,twocolumn]{revtex4-1}

\usepackage{bm,hyphenat,xspace}
\usepackage{graphicx,epsfig}
\usepackage{color}

\newcommand {\mbf}[1]{{\mathbf{#1}}}

\newcommand{\cm}{\mathrm{c\!\:\!.m\!\:\!.}}
\newcommand{\A}[2]{{}^{#1}\mathrm{#2}}
\newcommand{\SF}{\mathrm{SF}}

\begin{document}

\title {
Core-excitation effects in three-body breakup reactions studied
using the Faddeev formalism
}

\author{A.~Deltuva}
\email{arnoldas.deltuva@tfai.vu.lt}
\affiliation
{Institute of Theoretical Physics and Astronomy,
Vilnius University, Saul\.etekio al. 3, LT-10257 Vilnius, Lithuania
}
\affiliation{
Institute for Theoretical Physics II, 
Ruhr-University Bochum, D-44870 Bochum, Germany}

\received{12 December, 2018}

\begin{abstract}
\begin{description}
\item[Background]
Previous studies of $(d,p)$ reactions in three-body
(proton, neutron, nuclear core) systems revealed a nontrivial
effect of the core excitation: the transfer cross section
cannot be factorized  into the spectroscopic factor and the single-particle
cross section obtained neglecting the core excitation. This observable,
up to a kinematic factor, is the angular distribution of the core
nucleus in the $(p,d)$ reaction.
\item[Purpose]
The study of the core excitation effect for the most closely related
observable in the $(p,pn)$ three-body breakup, i.e., the core
angular distribution, is aimed in the present work.
\item[Methods]
Breakup of the one-neutron halo nucleus in the collision with the 
proton is described using three-body Faddeev-type equations extended
to include the excitation of the nuclear core.
The integral equations for transition operators are solved in the
momentum-space partial-wave representation.
\item[Results]
Breakup of $\A{11}{Be}$ nucleus as well as of model $A=11$ $p$-wave 
nuclei is studied at beam energies of 30, 60, and 200 MeV per nucleon.
Angular and momentum distributions for the $\A{10}{Be}$ core in
ground and excited states is calculated. In sharp contrast to
$(p,d)$ reactions, the differential cross section in most cases 
factorizes quite well into the spectroscopic factor and the single-particle
cross section.
\item[Conclusions]
Due to different reaction mechanisms the core excitation effect in the
breakup is very different from transfer reactions. A commonly accepted
approach to evaluate the cross section, i.e.,
the rescaling of single-particle model results
by the corresponding spectroscopic factor, 
appears to be reliable for breakup though it fails
in general for transfer reactions.
\end{description}
\end{abstract}

 \maketitle

\section{Introduction}

Excitation of the  core of the nucleus in few-cluster nuclear  reactions
is expected to be an important dynamic ingredient. Its effect has been studied
recently in three-body breakup reactions using the 
distorted-wave impulse approximation (DWIA) \cite{moro:12a,moro:12b} 
and extended continuum discretized coupled channels  (XCDCC) method 
\cite{summers:07a,diego:14a,diego:17a},
while for transfer reactions $(d,p)$ and $(p,d)$ exact Faddeev-type
equations in the extended Hilbert space have been employed
\cite{deltuva:13d}.
The latter works demonstrated that, in contrast to a widely accepted
assumption, the core excitation (CX) effect in transfer reactions
in general cannot be simulated by the spectroscopic factor (SF) only,
i.e., by the rescaling of the cross section obtained in the so-called
single-particle (SP) model that neglects the CX. The CX effect
due to rotational or vibrational quadrupole excitation becomes
most evident at energies around 50 MeV in the center-of-mass (c.m.)
system, and depends strongly on the angular momentum transfer $\ell$,
suppressing the forward-angle differential cross section for
$\ell=0$ and enhancing for $\ell=2$ \cite{deltuva:17b}.
Since those transfer calculations included breakup to all orders
and the breakup operator in the Faddeev formalism is obtained from
half-shell elastic and transfer operators, it is interesting and important
to investigate the CX effect for $(p,pn)$ breakup reactions.
The present work focuses on angular and momentum distributions of the core,
the observables that are experimentally measurable but are difficult to
 achieve the convergence in the XCDCC approach,
as found in the previous benchmark comparison of CDCC and Faddeev
calculations \cite{deltuva:07d}.

Section II presents the three-body Faddeev formalism for breakup reactions
including the CX; the unit convention $\hbar = c = 1$ is adopted. 
 Analysis of CX effects using several  $\A{11}{Be}$-like
model nuclei as examples is given in  Sec. III.
Section IV contains summary and conclusions.

\section{Theory}

I consider a three-particle system 
consisting of a proton ($p$), neutron ($n$),
and nuclear core ($C$). Odd-man-out notation is used where the channel of
a two-particle pair, the third one being a spectator, is labeled according
to the spectator and indicated by Greek subscripts.
When interacting with nucleons, the core can be excited or deexcited.
Ground (g) and excited (x) states are considered simultaneously 
in the extended Hilbert space \cite{deltuva:13d}
whose sectors are coupled by the interaction; the
respective components of the operators
are indicated by Latin superscripts. 
 Faddeev equations \cite{faddeev:60a} for transition operators $U_{\beta \alpha}$ 
in the version proposed
by  Alt, Grassberger, and Sandhas (AGS)
 \cite{alt:67a} were  formulated  in Ref.~\cite{deltuva:13d}
for  the Hilbert space with several sectors to enable the inclusion 
of the CX, i.e.,
\begin{equation}  \label{eq:Uba}
U_{\beta \alpha}^{ba}  = \bar{\delta}_{\beta\alpha} \, \delta_{ba} {G^{a}_{0}}^{-1}  +
\sum_{\gamma=p,n,C} \, \sum_{c=g,x}   \bar{\delta}_{\beta \gamma} \, T_{\gamma}^{bc}  \,
G_{0}^{c} U_{\gamma \alpha}^{ca}.
\end{equation}
Here $\bar{\delta}_{\beta\alpha} = 1 - \delta_{\beta\alpha}$,
$E$ is the energy in the c.m. frame,   and
 $G_0^{a} = (E+i0-\delta_{ax}\Delta m_C - K)^{-1}$ is
the free resolvent that does not couple Hilbert sectors,
but beside the internal-motion kinetic energy operator $K$ contains
also the contribution of the excitation energy $\Delta m_C$. 
For each pair $\alpha$ the potential $v_{\alpha}^{ba}$ leads to
the respective two-particle transition matrix
\begin{equation}  \label{eq:Tg}
T_{\alpha}^{ba} =  v_{\alpha}^{ba} +\sum_{c=g,x} 
v_{\alpha}^{bc} G_0^{c} T_{\alpha}^{ca}
\end{equation}
to be inserted into three-particle equations (\ref{eq:Uba}).
Elastic ($\alpha =\beta$) and transfer ($\alpha\neq \beta = p,n,C$)
operators were calculated in 
Refs.~\cite{deltuva:13d,deltuva:17b}. In this work I focus
on the breakup whose operator is  obtained from
Eqs.~(\ref{eq:Uba}) with $\beta=0$, i.e.,
\begin{equation}  \label{eq:U0a}
U_{0 \alpha}^{ba}  =  \delta_{ba} {G^{a}_{0}}^{-1}  +
\sum_{\gamma=p,n,C} \,  \sum_{c=g,x}   \, T_{\gamma}^{bc}  \,
G_{0}^{c} U_{\gamma \alpha}^{ca}.
\end{equation}
 Thus, formally it does not require a new 
solution of integral equations but is given as a quadrature involving
 elastic and transfer operators $U_{\gamma \alpha}^{ca}$.

The physical breakup amplitudes are determined by the on-shell matrix elements
of $U_{0 \alpha}^{ba}$ taken between initial two-cluster and final three-cluster
channel states. For a proton impinging with the relative 
momentum $\mbf{q}_p$
on a one-neutron halo nucleus the
initial state $|\Phi_p (\mbf{q}_p) \rangle = 
|\Phi_p^g(\mbf{q}_p) \rangle + |\Phi_p^x(\mbf{q}_p) \rangle$ 
has coupled ground- and excited-state core components.
In the final three-cluster channel 
$|\Phi_0^b (\mbf{p}'_\beta,\mbf{q}'_\beta) \rangle$  one
can separate ground ($b=g$) and excited ($b=x$) states of the core,
while the relative pair and spectator momenta $\mbf{p}'_\beta$ 
and $\mbf{q}'_\beta$ can be given in any of the three Jacobi sets.
Thus, the amplitude for the breakup of a halo nucleus leading
to its core state $b$ is given by
\begin{equation} 
\mathcal{T}^b_p(\mbf{p}'_\beta,\mbf{q}'_\beta;\mbf{q}_p) =
\sum_{a=g,x} \langle \Phi_0^b (\mbf{p}'_\beta,\mbf{q}'_\beta)|
U_{0 p}^{ba}  |\Phi_p^a (\mbf{q}_p) \rangle.
\end{equation}

Starting from the standard expression for the differential 
three-cluster breakup cross section 
\begin{gather}  \label{eq:d6s}
\begin{split}
d^6\sigma = {} & {} (2\pi)^4 \, \frac{M_p}{q_p} \,
\delta \left(E-\frac{{p'_\beta}^2}{2\mu_\beta} - \frac{{q'_\beta}^2}{2M_\beta} \right) \, \\
& \times |\mathcal{T}^b_p(\mbf{p}'_\beta,\mbf{q}'_\beta;\mbf{q}_p)|^2 \,
d^3\mbf{p}'_\beta d^3\mbf{q}'_\beta
\end{split}
\end{gather}
the semi-inclusive differential cross section 
for detecting only the particle $\beta$ is
\begin{gather}  \label{eq:d3s}
\frac{d^3\sigma}{d^3\mbf{q}'_\beta}
= (2\pi)^4 \, \frac{M_p}{q_p} \, \mu_\beta {p'_\beta} \, 
\int d^2\mbf{\hat{p}}'_\beta \, 
|\mathcal{T}^b_p(\mbf{p}'_\beta,\mbf{q}'_\beta;\mbf{q}_p)|^2.
\end{gather}
Here the magnitude of the relative pair momentum ${p'_\beta}$
is determined by the energy conservation that is reflected by the
$\delta$-function in Eq.~(\ref{eq:d6s}) 
with $\mu_\beta$ ($M_\beta$) being the pair (spectator) reduced mass.
Further integration of the threefold differential cross section 
(\ref{eq:d3s}) with respect to the magnitude or specific Cartesian
components of $\mbf{q}'_\beta$ leads to the angular 
$d^2\sigma/d^2\hat{\mbf{q}}'_\beta \equiv d\sigma/d\Omega_\beta$
or momentum $d\sigma/dq'_{\beta,i}$ distributions
for the particle $\beta$ in the c.m. frame.

The solution of the system of 
 integral equations (\ref{eq:Uba}) and the calculation of the quadrature
(\ref{eq:U0a}) is performed  in the momentum-space 
partial-wave framework, with a subsequent transformation to the
plane-wave representation used in Eq.~(\ref{eq:d3s}).
Partial waves with orbital momenta
$L_\alpha$ up to 5, 5, and 8 for the neutron-core, proton-neutron, and
proton-core pair, respectively, are included,
 the total angular momentum $J$ taking values up to 55.
With these truncations the results for semi-inclusive 
differential cross sections appear to be converged within 5\%,
while the difference between the CX and SP results, i.e., the CX 
effect, is converged even better. Thus, the achieved convergence 
is fully sufficient for a reliable study of the CX effect.
Further details on the implementation of three-body reaction
calculations including the CX  can be found in 
Refs.~\cite{deltuva:13d,deltuva:17b}.

\section{Results}

\begin{figure*}[!]
\begin{center}
\includegraphics[scale=0.7]{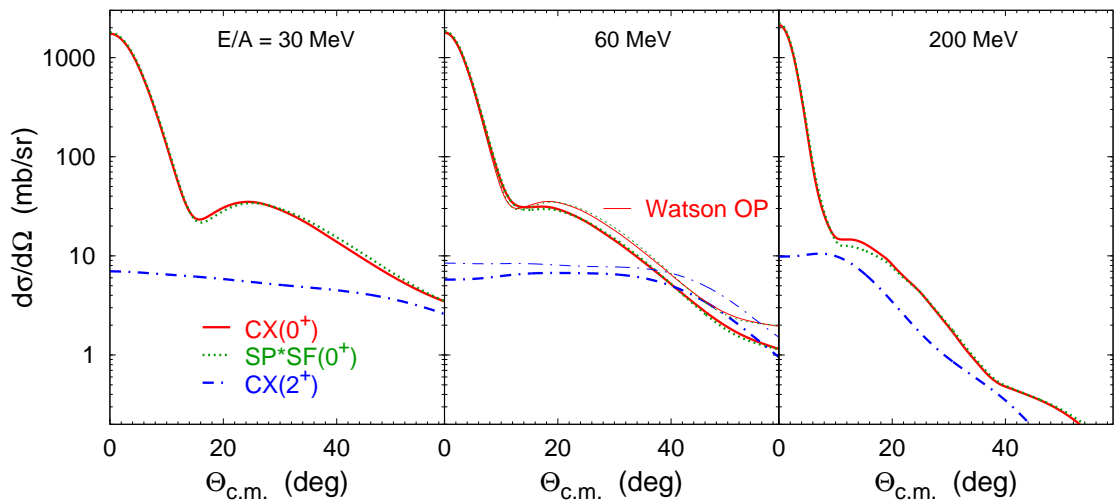}
\end{center}
\caption{\label{fig:sa}
Semi-inclusive differential cross sections for 
$\A{11}{Be}(p,pn)\A{10}{Be}$ reactions
at $E/A = 30$, 60,  and 200 MeV 
as functions of the  $\A{10}{Be}$ core c.m. scattering angle.
Predictions of models with CX, displayed by solid (dashed-dotted) curves for 
$0^+$ ($2^+$) final states of the $\A{10}{Be}$ core, 
and without CX, the latter  rescaled by $\SF(0^+) = 0.854$
and displayed by dotted curves, are compared.
In addition to the results based on the KD optical potential,
the ones based on the Watson parametrization are shown at 60 MeV
as thin curves.}
\end{figure*}

Previous DWIA \cite{moro:12a,moro:12b} 
and XCDCC  \cite{summers:07a,diego:14a,diego:17a} studies of 
 breakup of one-neutron halo nuclei found an 
 important CX effect for the resonant breakup  when the resonance 
has a significant component with excited core.
The present work focuses  
on the study of the global CX effect in angular and momentum
distributions in $(p,pn)$ reactions
and its comparison with the CX effect in $(p,d)$ reactions
\cite{deltuva:13d,deltuva:16c}.
Therefore the example three-body system  $p+n+\A{10}{Be}$
used in this investigation is taken from  the corresponding
$(d,p)$ study \cite{deltuva:16c}.
For the core nucleus $\A{10}{Be}$ the ground  $0^+$ and excited $2^+$ states
are considered with the excitation energy $\Delta m_C = 3.368$ MeV.

The dynamics input for the scattering equations of the previous
section is determined by three pairwise multicomponent potentials 
$v_{\alpha}^{ba}$. For the $n$-$p$ subsystem there is a number 
of realistic interaction models such as the 
charge-dependent Bonn (CD Bonn) potential \cite{machleidt:01a}, 
that will be used also in the present work. 
This is an important improvement compared to the previous
XCDCC calculation \cite{diego:17a} of angular and energy distributions
that used simple Gaussian $n$-$p$ potential; the failure of the Gaussian
potential in reproducing the breakup results of realistic potentials
was demonstrated without CX in Ref.~\cite{cravo:10a}.
The neutron-core binding potentials for the spin/parity $j^\pi = \frac12^+$ state of 
 $\A{11}{Be}$ with the neutron separation energy $S_n = 0.504$ MeV
is taken over from Ref.~\cite{deltuva:16c}, whereas Watson \cite{watson}
and Koning and Delaroche (KD) \cite{koning} optical potentials are
used for proton-core and neutron-core interaction in remaining partial waves.
The proton-core Coulomb interaction is included as well using the method
of screening and renormalization \cite{alt:80a}.
The CX is included via the standard rotational quadrupole deformation 
of the underlying potentials \cite{tamura:cex,nunes:96a}
with the deformation parameter 
 $\beta_2 = 0.67$  and the deformation length  $\delta_2 = 1.664$ fm,
while the subtraction technique \cite{deltuva:15b} ensures the two-body
on-shell equivalence of the models with and without CX at the given energy.
When the reaction is initiated by the beam of $E/A$
energy per nucleon, the energy-dependent
optical potential parameters are taken at fixed energy values of
$E/A$ and $\frac12(E/A)$ for $p$-$C$ and $n$-$C$ interactions, respectively.
The Watson parametrization was designed for light p-shell nuclei but 
is constrained by the data up to 50 MeV. In contrast, 
the use of the KD parametrization for $\A{10}{Be}$ can be criticized
as it was fitted to the data for $A\geq 24$ nuclei, but it has an advantage 
of applicability in a broader energy interval extending up to 200 MeV.
Comparison of predictions based on Watson and KD potentials will estimate
the associated uncertainty.

Previous studies of neutron transfer reactions 
\cite{deltuva:13d,deltuva:15b,deltuva:17b,deltuva:16c,gomez:17a}
revealed a nontrivial CX effect in the $(d,p)$ differential cross section.
This observable, up to kinematic and spin factors, coincides with the
angular distribution of the core in the three-body c.m. frame 
in  $(p,d)$ transfer reactions.
It is therefore interesting to investigate  the most
closely related observable in the $(p,pn)$ breakup, i.e., the angular core 
cross section $d\sigma/d\Omega_C$ in the c.m. frame.
The difference to the $(p,d)$ cross section is that the $n$-$p$ pair is not
bound but can be in any continuum state up to the allowed energy;
an integration over those states is needed as explained in the previous section.
For brevity, the subscript denoting the core in the following will be omitted,
both for the angular $d\sigma/d\Omega$ and momentum 
$d\sigma/dq_i$ distributions.

The SP and CX differential cross sections 
 $d\sigma/d\Omega$ as functions of the core c.m. scattering angle $\Theta_\cm$
at beam energies of 30, 60, and 200 MeV per nucleon
are compared in Fig.~\ref{fig:sa}. $E/A = 60$ MeV roughly corresponds
to the maximal CX effect in transfer reactions \cite{deltuva:16c},
while 200 MeV should represent the region of relatively high energy
with ongoing experimental activities.
For a better comparison the SP results are renormalized by the
$\SF(0^+) = 0.854$ for the $\A{10}{Be}(0^+)$ component in the 
$\A{11}{Be}(\frac12^+)$ bound state.
These rescaled SP results
simulate well the $\A{10}{Be}(0^+)$ angular distribution in the model
including the CX explicitly.  Although the shape of the observable
changes with the energy, one can see qualitatively the same agreement 
between these two types of results
 in all considered cases, also when replacing the KD optical potential by that
of Watson.
In fact, the  forward peak of the $\A{10}{Be}(0^+)$ angular distribution 
appears to 
be quite insensitive to the choice of the optical potential, in contrast
to larger angles and 
the $\A{10}{Be}(2^+)$ angular distribution where the Watson parametrization
leads to a larger differential cross section compared to KD, much like
for the $(p,d)$ transfer in Ref.~\cite{deltuva:15b}.

\begin{figure}[!]
\begin{center}
\includegraphics[scale=0.8]{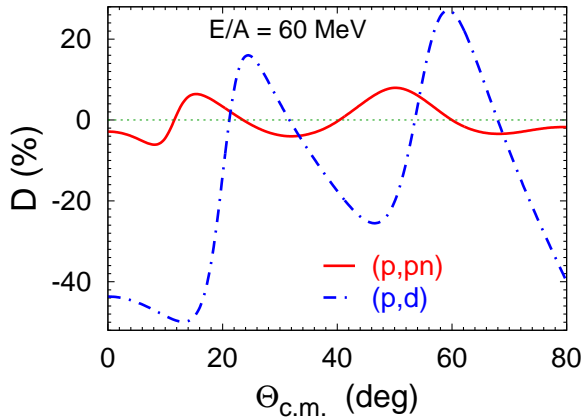}
\end{center}
\caption{\label{fig:sr}
Core excitation effect for the core angular distribution
in breakup (solid curve) and transfer (dashed-dotted curve) reactions,
resulting from the proton-$\A{11}{Be}$ collision
at $E/A = 60$ MeV.}
\end{figure}

Such a behavior is in  sharp contrast with transfer reactions
\cite{deltuva:13d,deltuva:15b,deltuva:17b,deltuva:16c,gomez:17a} 
where  $d\sigma/d\Omega(0^+)$ obtained including the CX
cannot be factorized into the SP differential cross section 
$(d\sigma/d\Omega)_{SP}$ and the associated SF. 
The size of this CX effect may be characterized  by the ratio
\begin{equation} \label{eq:R}
R = \frac{d\sigma/d\Omega(0^+)}{\SF(0^+)\cdot (d\sigma/d\Omega)_{SP}}
\end{equation}
or its deviation from unity  $D = (R-1)\times 100\%$.
In Fig.~\ref{fig:sr} the latter is compared for breakup and transfer reactions
at $E/A = 60$ MeV, i.e., in the energy region showing the largest CX effect 
for transfer \cite{deltuva:16c}. 
Despite some differences in employed optical potentials, 
the  CX effect  in the $(p,d)$ reaction shown in Fig.~\ref{fig:sr}
 is in agreement with Ref.~\cite{deltuva:16c}. It is most
sizable at  forward angles,  where the cross section peaks,
reaching almost 50\% in magnitude. 
On the contrary, the same characteristic quantity $D$ for the breakup 
stays well below 10\% in magnitude, 
as can be expected from Fig.~\ref{fig:sa}.

\begin{figure}[!]
\begin{center}
\includegraphics[scale=0.56]{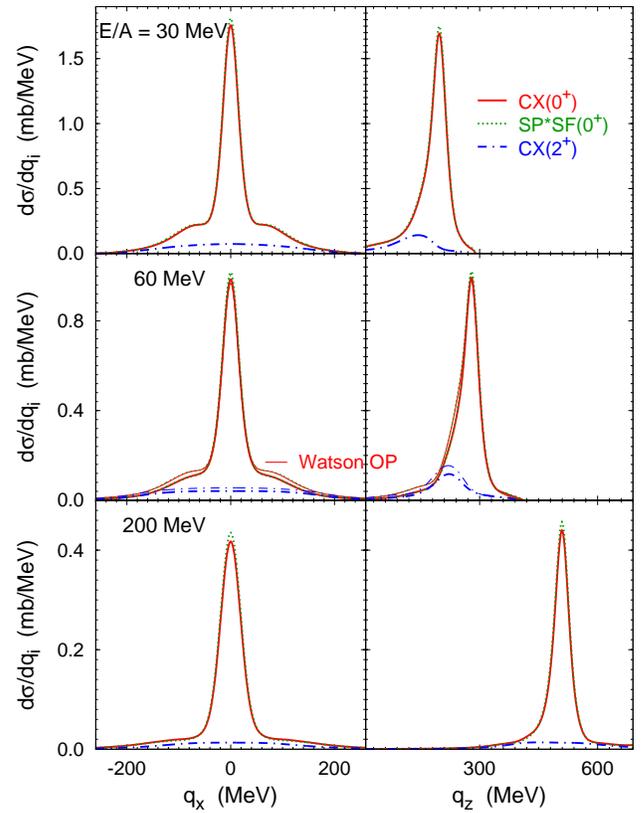}
\end{center}
\caption{\label{fig:sm}
Transverse (left) and longitudinal (right) core momentum
distributions for 
$\A{11}{Be}(p,pn)\A{10}{Be}$ reactions
at $E/A = 30$, 60,  and 200 MeV.
Curves are as in Fig.~\ref{fig:sa}.}
\end{figure}

The transverse and longitudinal core momentum distributions 
$d\sigma/dq_i$ at the same beam energies  are
shown in Fig.~\ref{fig:sm}. Again, SP results multiplied by the
$\SF(0^+)$ simulate well  $\A{10}{Be}(0^+)$ momentum distributions 
including the CX, with small differences of about 3-4\%
seen at the peaks.
The optical potential sensitivity is studied at  $E/A = 60$ MeV where
the Watson potential predictions are similar to those of KD 
for $\A{10}{Be}(0^+)$ at the peaks 
but are higher by about 20 - 30\% at the shoulders and for
$\A{10}{Be}(2^+)$ momentum distributions.
The latter is broader and much smaller in the absolute value,
as can be expected due to a small  $\SF(2^+) = 0.146$, 
an effectively larger binding $S_n + \Delta m_C$ and 
$d$-wave excited core component in $\A{11}{Be}$.

The study of transfer reactions \cite{deltuva:15b} revealed also the dependence
of the CX effect on the internal orbital momentum  of the bound state,
$p$-wave systems exhibiting weaker CX effect of  opposite sign as compared
to $s$-wave. 
For this reason I consider  two fictitious $p$-wave model 
 nuclei of mass $A=11$ with the core of $\A{10}{Be}$ but of
spin/parity $\frac12^-$ with neutron separation energies 
$S_n = 0.5$ and 5.0 MeV. 
The first value nearly equals the $s$-wave case of the above 
$\A{11}{Be}(\frac12^+)$ breakup study and represents very weakly bound system,
 while the latter is a more typical value for $p$-shell nuclei.
Although both model nuclei differ from the physical $\A{11}{Be}(\frac12^-)$ excited state
with $S_n = 0.184$ MeV, for brevity they will be referred to as $\A{11}{Be}(\frac12^-)$.
The parameters of the respective neutron-core binding potentials  
in the $\frac12^-$ partial wave are taken over from Ref.~\cite{deltuva:16c},
except for the central Woods-Saxon strength adjusted to the desired binding
energy. Otherwise, optical potentials as in the above study of
$\A{11}{Be}(\frac12^+)$ breakup are used.

\begin{figure}[!]
\begin{center}
\includegraphics[scale=0.56]{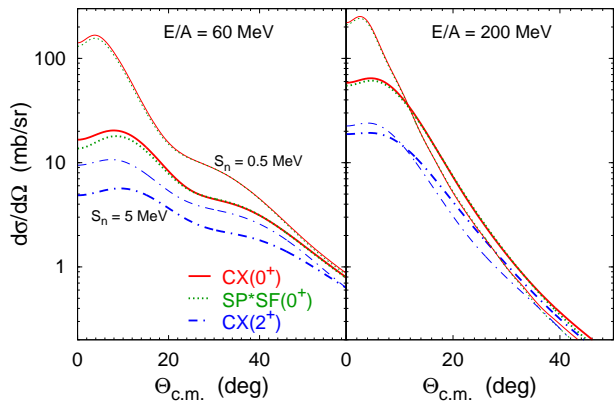}
\end{center}
\caption{\label{fig:pa}
Semi-inclusive differential cross sections for 
the breakup of $p$-wave model nuclei 
as functions of the  core c.m. scattering angle at $E/A = 60$  and 200 MeV.
Predictions of models with CX, displayed by solid (dashed-dotted) curves for 
$0^+$ ($2^+$) final states of the  core, are compared with  rescaled results of the SP model.
Thin (thick) curves correspond to the neutron separation energy $S_n = 0.5$ MeV (5.0 MeV)
with $\SF(0^+)$ being 0.771 (0.731).  KD optical potential is used.}
\end{figure}

The angular  distributions of the
$\A{10}{Be}$ core resulting from the $p$-$\A{11}{Be}(\frac12^-)$ breakup
at  $E/A = 60$  and 200 MeV are shown in Fig.~\ref{fig:pa}. 
Again, SP predictions renormalized by the respective 
$\SF(0^+) = 0.771$ (0.731) for the $\A{11}{Be}(\frac12^-)$ binding of
0.5 (5.0) MeV reproduce reasonably well the observables including CX,
except  that a more sizable deviation of 20\% 
is observed at $\Theta_\cm =0^\circ$ for the breakup of the more
tightly bound nucleus at the lower beam energy. Such a combination can be 
associated with larger rescattering contributions, that enhance the importance 
of nucleon-core interactions and, as a consequence, the CX effect.

\begin{figure}[!]
\begin{center}
\includegraphics[scale=0.56]{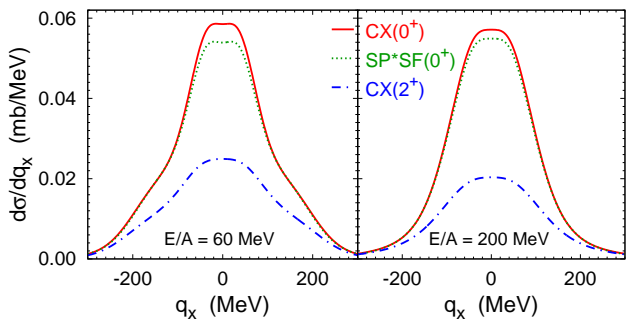}
\end{center}
\caption{\label{fig:pm}
Transverse  core momentum distributions for the breakup of $p$-wave model nucleus
with 5.0 MeV binding energy 
in the collision with proton at $E/A = 60$  and 200 MeV.
Curves are as in Fig.~\ref{fig:pa}.}
\end{figure}

Since transverse and longitudinal core momentum distributions show very similar 
CX effect, I present in  Fig.~\ref{fig:pm}  only the former for a more 
tightly bound model nucleus with $S_n = 5.0$ MeV. At a lower beam energy
 $E/A = 60$ MeV 
in the peak of the $0^+$ core transverse momentum distribution 
there is a clear difference of about 8.5\% between the results including CX 
and those of the SP model rescaled by $\SF(0^+) = 0.731$.
That difference is however reduced to 4.0\%
at a higher beam energy  $E/A = 200$ MeV where the rescattering is less important.
In the case of $S_n = 0.5$ MeV, not shown here, the corresponding difference at the
peaks of $0^+$ core  momentum distributions amount to 5.9\% (3.5\%) at 
 $E/A = 60$ MeV (200 MeV).

Finally, Table \ref{tab} compares also predictions for
the integrated three-body breakup cross section 
$\sigma$ and its components with the
core being in a given state.
The CX effect for the ground-state core cross section $\sigma(0^+)$
is characterized by the parameter $\bar{D}$ defined as $D$ but with
the differential cross sections in Eq.~(\ref{eq:R}) replaced by the 
integrated ones. In all considered cases its magnitude remains below 5\%,
showing quite  weak sensitivity to the beam and binding energies
and optical potential.
$\bar{D}$ slightly depends on the orbital angular momentum of $\A{11}{Be}$,
being negative for the breakup of $\A{11}{Be}(\frac12^+)$
but positive for $\A{11}{Be}(\frac12^-)$.
The CX effect on the total cross section including all states is quantified
by $\bar{D}_\Sigma = (\sigma/\sigma_{SP}-1)\times 100\%$ that
is listed in the Table \ref{tab}  as well. This quantity shows more 
 sensitivity to the optical potential, beam energy, and $\A{11}{Be}$
bound state properties, mainly caused by the $\sigma(2^+)$ contribution.

\begin{table} [!]
\caption{\label{tab}
Integrated three-cluster breakup cross sections 
(in millibarns) for proton-$\A{11}{Be}$
collisions at given beam energies (in MeV), calculated using various models for
$\A{11}{Be}$, characterized by the neutron separation energy (in MeV) and spin/parity. The CX effect for $0^+$ and total cross sections
(in percents) is given in the two last columns.
KD optical potential was used, except for the 3rd line results
derived from the Watson parametrization.
}
\begin{ruledtabular}
\begin{tabular}{*{8}{r}}
$S_n(j^\pi)$ & $E/A$ & 
$\sigma(0^+)$ & $\sigma(2^+)$ & 
$\sigma$ & $\sigma_{SP}$ & $\bar{D}$ &  $\bar{D}_\Sigma$   \\ \hline
 $0.504(\frac12^+)$ & 30 & 114.0 & 20.3 & 134.3 & 138.2 & -3.4 & -2.8 \\
 $0.504(\frac12^+)$ & 60 & 67.1 & 14.9 & 82.0 & 80.4 & -2.3 & 2.0 \\
 $0.504(\frac12^+)$ & (W)60 & 72.5 & 20.8 & 93.3 & 87.7 & -3.2 & 6.4 \\ 
 $0.504(\frac12^+)$ &200 & 27.7 & 4.4 & 32.1 & 33.5   & -3.2 & -4.2 \\
 $0.500(\frac12^-)$ & 60 & 35.9 & 10.3 & 46.2 & 44.5 & 4.6 & 3.8 \\
 $0.500(\frac12^-)$ &200 & 19.3 & 5.4 & 24.7 & 24.2 & 3.4 & 2.0 \\
 $5.000(\frac12^-)$ & 60 & 14.8 & 7.5 & 22.3 & 19.3 &  4.9 & 15.5 \\
 $5.000(\frac12^-)$ &200 & 13.6 & 5.9 & 19.5 & 18.0 & 3.4 & 8.3 \\
\end{tabular}
\end{ruledtabular}
\end{table}

\section{Discussion and conclusions}

The three-body system  of proton, neutron, and nuclear core
was treated in the extended Faddeev-type formalism allowing
the excitation of the nuclear core.
Integral equations for three-body transition operators were solved
 in the momentum-space partial-wave basis.

Breakup of the one-neutron halo nucleus  $\A{11}{Be}$ in the collision 
with the proton was considered. 
Angular and transverse and longitudinal momentum distributions 
of the $\A{10}{Be}$ core were calculated at beam energies of 30, 60, 
and 200 MeV/nucleon. 
For the core detected in its ground state $0^+$ the results appear to be
quite insensitive to the choice of the nucleon-core optical potential.
The single-particle calculations that neglect the CX, renormalized by the  
corresponding spectroscopic factor $\SF(0^+)$ 
of the $\A{11}{Be}$ bound state, simulate quite well the 
differential cross sections obtained including the CX.

In order to study the dependence on the internal orbital angular momentum
 of the bound state, two fictitious $\A{11}{Be}$-like 
$p$-wave nuclei with neutron separation energies of 0.5 and 5.0 MeV
were considered as well. Despite different shapes of 
differential cross sections, the CX effect turns out to be quite similar,
i.e., the factorization of the cross section for the
$\A{10}{Be}(0^+)$ state into the SP cross section and the respective
$\SF(0^+)$ remains quite a good approximation.  Some deviations are
seen mostly at the peaks of distributions, with the most sizable one appearing
at the lower beam energy and larger binding energy, where one may
expect larger nucleon-core rescattering contributions, 
enhancing also the importance of the CX.

Integrated three-body breakup cross sections were also studied, 
leading to a similar 
conclusion --- the CX effect for the core ground-state
cross section largely consists in the 
renormalization of the SP cross section by the corresponding SF.
The CX effect on the total cross section depends more strongly 
on dynamic details.

Thus, CX effects in coupled breakup and neutron transfer reactions 
turn out to be
very different, a probable reason being different reaction mechanisms. 
Breakup, especially at higher energies, is dominated by the
neutron-proton quasi free scattering (QFS) where the proton knocks out the 
neutron  from the initial nucleus while subsequent interactions
between nucleons and the remaining nuclear core are responsible
for the distortion, typically reducing the cross section. 
Neglecting this distortion, i.e., in the plane-wave impulse approximation,
the differential cross section is proportional to the square of the
momentum-space bound-state wave function.
A consequence of this reaction mechanism is a substantial sensitivity to 
the neutron-core  bound-state wave function, 
surviving also after the distortion. In fact, the differences
in Figs.~\ref{fig:sa} - \ref{fig:pm} between CX($0^+$) and rescaled SP 
predictions to some extent may be caused by small differences
in the shapes of the corresponding $\A{11}{Be}$ wave function
components.
Nevertheless, the similarity of the nuclear wave functions
leads to the observed scaling of 
breakup cross sections calculated with and without the CX.
In contrast, the transfer  reaction mechanism involves high-order
rescattering between all three involved particles, smearing out the
sensitivity to the details of the wave function and enhancing the
importance of nucleon-core interactions, thereby also the dynamic 
CX effect.

\begin{acknowledgments}
I acknowledge the discussions with R.~Crespo and A.~M.~Moro, and 
the support  by the Alexander von Humboldt Foundation
under Grant No. LTU-1185721-HFST-E and by the
Funda\c c\~ao Ci\^encia e Tecnologia (FCT) of Portugal
under  Contract No. PTD/FIS-NUC/2240/2014.
\end{acknowledgments}


\end{document}